\documentclass[aps,physrev,reprint,groupedaddress]{revtex4-2}
\usepackage[utf8]{inputenc}
\usepackage{booktabs}
\usepackage{amsmath}
\usepackage{graphicx}

\makeatletter

\DeclareFontEncoding{LGR}{}{}

\providecommand{\tabularnewline}{\\}

\usepackage{braket}

\makeatother

\begin{document}
\title{Excited States from Quasiparticle Hamiltonian Based on Density Functional
Theory}
\author{Yang Shen}
\thanks{These authors contribute equally to this work.}
\author{Yichen Fan}
\thanks{These authors contribute equally to this work.}
\affiliation{Department of Chemistry, Duke University, Durham, NC 27708}
\author{Weitao Yang}
\email{weitao.yang@duke.edu}
\affiliation{Department of Chemistry and Department of Physics, Duke University,
Durham, NC 27708}
\date{\today}
\begin{abstract}
Recent advances in occupancy extrapolation (OE) show that potential of orbital-occupation-based energy functions can describe electronic excitations. Here, the OE method in the particle-hole channel is extended to an effective quasiparticle Hamiltonian, enabling a multi-configurational description beyond single-determinant OE and $\Delta$SCF. The method performs comparably to the Bethe–Salpeter equation for valence singlet and charge-transfer excitations, and better for valence triplet and Rydberg states, supporting its accuracy and broad applicability.
\end{abstract}
\maketitle
The determination of excited state energy and transition properties
is central in both experimental characterization and theoretical prediction
of photoabsorption spectra and photodynamics. Among widely used methods,
wavefunction-based approaches such as equation-of-motion coupled cluster
(EOM-CC)~\citep{stanton_equation_1993} and algebraic diagrammatic
construction (ADC)~\citep{schirmer_new_1983,schirmer_non-dyson_1998,dreuw_algebraic_2015,banerjee_algebraic_2023}
are accurate for small systems but often regarded as too computationally
demanding for large molecules and condensed-phase systems. Time-dependent
density functional theory (TDDFT)~\citep{runge_density-functional_1984,doi:10.1142/9789812830586,bauernschmitt_treatment_1996,dreuw_single-reference_2005,ullrich_time-dependent_2012,hu_accelerating_2020}
offers more favorable scaling, ($O(N^{3}-N^{4})$), but its accuracy
can be limited by the deficiencies of the density functional approximation
(DFA). For intermolecular charge-transfer (CT) excitations, the absense
of 100\% Hartree-Fock exchange at long range leads to incorrect asymptotic
behavior of the CT excitation energies\citep{dreuw_failure_2004,dreuw_single-reference_2005}.
For diffuse Rydberg states, the rapid exponential decay of exchange
correlation potential produces large errors in excitation energy\citep{della_sala_excitation_2003,cheng_rydberg_2008}.

As an alternative, the Bethe-Salpeter equation (BSE), based on the
many-body perturbation theory, or Green's function theory, \citep{salpeter_relativistic_1951,rohlfing_electron-hole_2000,onida_electronic_2002,bruneval_systematic_2015,blase_bethesalpeter_2018,krause_implementation_2017,liu_all-electron_2020,loos_dynamical_2020,zhou_all-electron_2025}
can adopt a full or partial self-consistency in quasiparticle energy
to alleviate the functional dependence\citep{jacquemin_benchmarking_2015,blase_bethesalpeter_2018}.
In contrast to TDDFT, BSE involves quasiparticles and describes the
effective particle-hole interaction through the self-energy derivatives\citep{onida_electronic_2002,martin_interacting_2013}.
In combination with GW approximation for the self energy\citep{hedin_new_1965,hybertsen_electron_1986,lischner_first-principles_2012,ren_resolution--identity_2012,van_setten_gw100_2015,jiang_gw_2016,golze_gw_2019,zhu_all-electron_2021},
the BSE formalism incorporates screened exact exchange into the effective
particle-hole interaction. The screened exact exchange yields correct
asymptotic behavior and significantly improves the accuracy for CT\citep{bhattacharya_bsegw_2024}
and Rydberg excitations\citep{jacquemin_is_2017}. However, the cost
of computing accurate quasiparticle energies may exceed the cost of
solving BSE\citep{golze_gw_2019,10.1063/5.0260895}. Even with exact
quasiparticle energies, BSE remains limited by the neglect of vertex
corrections.

Another widely used class of excited-state methods is the $\Delta$
Self-Consistent Field ($\Delta$SCF) \citep{slater_statistical_1970,SLATER19721,bagus_singlettriplet_1975,ziegler_calculation_1977,RevModPhys.61.689,gilbert_self-consistent_2008,cheng_rydberg_2008,besley_self-consistent-field_2009,doi:10.1021/acs.jctc.6b01161,hait_accurate_2020,hait_orbital_2021,vandaele_scf_2022},
whose theoretical foundation has been established only recently~\citep{yang_foundation_2024}.
By promoting electrons from occupied to virtual orbitals, $\Delta$SCF
optimizes the total energy of a targeted excited state directly, with
a computational cost comparable to ground-state DFT. Nevertheless,
as a single-determinantal method, among other multi-determinant related
issues, $\Delta$SCF is subject to severe spin contamination for open-shell
states\citep{seidu_applications_2015,ye_-scf_2017,hait_accurate_2020,hait_orbital_2021},
requiring additional spin purification treatment\citep{bagus_singlettriplet_1975,ziegler_calculation_1977}.
In addition, excited-state SCF calculations can often encounter convergence
difficulties\citep{vandaele_scf_2022}.

To calculate the excitation energies without any additional $\Delta$SCF
calculation of excited state, Occupancy Extrapolation (OE) formulates the excitation
energy as a function of orbital occupation numbers $\{n\}$, around
a reference system, which is usually taken as a ground state\citep{fan_OE_2026}.
Connection and comparison of the OE method with the Landau Fermi liquid
theory\citep{lfliquid1, lfliquid2, baym_landau_2008, negele_quantum_2018} and Slater's work on transition
state and related approaches\citep{slater_statistical_1970, SLATER19721} has been described
in details\citep{fan_OE_2026}. We now briefly describe the OE method. The total
energy expression is given in equation \ref{eq:OE_tot}, truncated
here at the second order in a Taylor expansion,

\begin{equation}
\begin{aligned}
E_{\text{tot}}(\{n_{p\sigma}\})\approx &E_{\text{tot}}^{\text{DFA}}+\sum_{p\sigma}\frac{\partial E_{\text{tot}}^{\text{DFA}}}{\partial n_{p\sigma}}|_{n=n^{0}}(n_{p\sigma}-n_{p\sigma}^{0}) 
+\frac{1}{2}\\&\sum_{p\sigma,q\tau}\frac{\partial^{2}E_{\text{tot}}^{\text{DFA}}}{\partial n_{p\sigma}\partial n_{q\tau}}|_{n=n^{0}}(n_{p\sigma}-n_{p\sigma}^{0})(n_{q\tau}-n_{q\tau}^{0}),\label{eq:OE_tot}
\end{aligned}
\end{equation}

\noindent in which $E_{\text{tot}}^{0}$ is the total energy of the reference
state, with orbital occupation $\{n^{0}_{p\sigma}\}$,and p($\sigma$), q($\tau$) are the indices
for canonical (spin-)orbitals. The first-order derivatives, $\epsilon_{p\sigma}$, are the eigenvalues
of the Kohn-Sham Hamiltonian, as stated in the Janak theorem\citep{janak_proof_1978}
for continuous functionals of electron density, or the eigenvalues
of the generalized Kohn-Sham Hamiltonian, following the work of Cohen, Mori-Sanchez and Yang\citep{cohen_fractional_2008} for continuous functionals
of the non-interacting reference one-electron density matrix, such
as hybrid functionals.The
second-order derivatives describe quasiparticles and interaction between quasiparticles.
The diagonal second derivatives were first derived by Yang \textit{et al.}\citep{yang_analytical_2012}, and the complete second derivatives are the following\citep{mei_exact_2021},

\begin{equation}
\begin{aligned}
\kappa_{p\sigma,q\tau}\equiv & W_{pp\sigma,qq\tau}\\ \equiv & \frac{\partial^{2}E_{\text{tot}}^{\text{DFA}}}{\partial n_{p\sigma}\partial n_{q\tau}}=(\psi_{p\sigma}\psi_{p\sigma}|K+K\chi K|\psi_{q\tau}\psi_{q\tau}).\label{eq:kappa}
\end{aligned}
\end{equation}
where K is the spin-resolved Hartree-exchange-correlation kernel,
defined as the functional derivative of the spin (G)KS Hamiltonian
w.r.t. the spin density matrix $\gamma_{s}$, and $\chi$
 is the spin density (matrix) linear response function, all of which
are 4-point quantities in general and can be calculated directly from
the ground state DFA calculations\citep{yang_analytical_2012}.
Therefore OE is a method for obtaining excited state energy without
any additional SCF calculations beyond the ground state calculation.
In addition to the total energies of excited states, the orbital energies
for the excited states can also be obtained as the derivatives of
the OE expansion in Eq. \ref{eq:orb_ene_ex} \citep{fan_OE_2026}.

\begin{equation}
    \epsilon^{*}_{p\sigma} = \epsilon_{p\sigma}+\sum_{q\tau} \kappa_{p\sigma, q\tau} \delta n_{q\tau}.\label{eq:orb_ene_ex}
\end{equation}

The connection to an $(N+1)$-electron single-determinant state, in
which a virtual canonical orbital $\psi_{a\sigma}$ is occupied, is
established by the quasi-particle energy $\epsilon_{a\sigma}^{\text{QP}}$.
In the OE framework, the quasiparticle energy is described by the
one-particle channel, i.e. all $\delta n_{p\sigma}=0$ except for
$\delta n_{a\sigma}=1$. Truncated at the second order, the quasiparticle
energy is approximated in the OE expansion, Eq. \ref{eq:OE_QP}, as the sum of the
DFA orbital energy and the generalized screened interaction energy,
$\kappa_{a\sigma,a\sigma}$ ($\kappa_{i\sigma,i\sigma}$),

\begin{equation}
\epsilon_{a\sigma}^{\text{QP}}\approx\Delta E(1p)=\epsilon_{a\sigma}+\frac{1}{2}\kappa_{a\sigma,a\sigma}.\label{eq:OE_QP}
\end{equation}

The last equality is also exactly the result of GSC2 (global scaling
correction with exact second-order corrections) approximation to the
quasiparticle energy, which performs very well for small
molecules\citep{mei_exact_2021}. Similarly the quasiparticle energy
corresponding to an $(N-1)$-electron state can be approximated as
$\epsilon_{i\sigma}^{\text{QP}}\approx\Delta E(1h)=\epsilon_{i\sigma}^{\text{DFA}}-\frac{1}{2}\kappa_{i\sigma,i\sigma}$,
also a result of GSC2 \citep{mei_exact_2021}. Beyond the quasiparticle
energy, OE is also applicable to particle-hole neutral excitations.
The neutral excitation energy of a single-determinant state, in which one electron is promoted from an occupied canonical orbital $\psi_{i\sigma}$ to a virtual canonical orbital $\psi_{a\tau}$, can be expressed as\citep{fan_OE_2026}

\begin{equation}
    \Delta E(1p1h)=\epsilon_{a\tau}^{\text{QP}}-\epsilon_{i\sigma}^{\text{QP}}-W_{ii\sigma,aa\tau}.\label{eq:eh_excitation_OE}
\end{equation}

\noindent which can be viewed as the energy difference of quasi-particle, quasi-hole and the attraction
between them. This picture is essentially similar to that of BSE@GW
for single-determinant triplets.

Nevertheless, many excited states, including open-shell singlets, are essentially multi-configurational,
requiring additional treatment. Therefore, a state-average assumption
is made so that the energy of the broken-symmetry (B.S.) state with $\delta n_{a\sigma}=1$
and $\delta n_{i\sigma}=-1$ is assumed to be the average of singlet
and triplet energies. Without spin-orbit coupling and an external magnetic
field, the 3-fold degeneracy of triplets is preserved by the SU(2)
symmetry, which allows the triplet excitation energy to be accessed
by the spin-pure $S_{\text{z}}=\pm1$ triplet states. This leads to
the spin purification method for singlet state energies\citep{bagus_singlettriplet_1975, ziegler_calculation_1977},

\begin{equation}
\Delta E_{\text{S}}=2\Delta E_{\text{B.S.}}-\Delta E_{\text{T},S_{\text{z}}=\pm1}, \label{eq:spin_purification}
\end{equation}

\noindent where $\Delta E_{\text{B.S.}}$ is excitation energy of the broken symmetry state and $\Delta E_{\text{T},S_{\text{z}}=\pm1}$ is the excitation energy of the single-determinant triplet. Based on the quasiparticle-quasihole attraction picture and spin purification
process, OE provides accurate estimates of neutral excitation energies,
spanning a wide range of valence, CT and Rydberg states\citep{fan_OE_2026}. 

However, a critical drawback is that
the multi-configurational nature of the particle-hole wavefunction is completely absent from the OE description, or its parent $\Delta$SCF method. It is this deficiency that we aim to overcome in our present work. Viewing the OE energy function as a functional
of the particle-hole wavefunction, $\Delta E[\Psi_{\text{ph}}(\mathbf{x}_{\text{p}},\mathbf{x}_{\text{h}})]$,
the domain of definition is restricted to $\Psi_{\text{ph}}(\mathbf{x_{\text{p}}},\mathbf{x}_{\text{h}})=\Psi_{\text{\ensuremath{i\sigma,a\tau}}}(\mathbf{x_{\text{p}}},\mathbf{x}_{\text{h}})=\psi_{i\sigma}(\mathbf{x_{\text{h}}})\psi_{a\tau}^{*}(\mathbf{x}_{\text{p}})$.
This is essentially a single-determinant constraint of the particle-hole
wavefunction. An multi-determinant extension of the domain
of definition can be made, so that $\Psi_{\text{ph}}(\mathbf{x_{\text{p}}},\mathbf{x_{\text{h}}})=\sum_{i\sigma,a\tau}X_{i\sigma,a\tau}\psi_{i\sigma}(\mathbf{x_{\text{h}}})\psi_{a\tau}^{*}(\mathbf{x}_{\text{p}})$.
The expansion coefficients, $X_{i\sigma,a\tau}$, satisfy the normalization
condition $\sum_{i\sigma,a\tau}X_{i\sigma,a\tau}^{*}X_{i\sigma,a\tau}=1$.
Accordingly, the energy functional of the particle-hole wavefunction
can be written as

\begin{equation}
\begin{aligned}
\Delta E[\Psi_{\text{ph}}(\mathbf{x_{\text{p}}},\mathbf{x_{\text{h}}})]=\langle\Psi_{\text{ph}}|\sum_{i\sigma,a\tau}[(\epsilon_{a\tau}^{\text{QP}}-\epsilon_{i\sigma}^{\text{QP}})|\Psi_{i\sigma,a\tau}\rangle\\\langle\Psi_{i\sigma,a\tau}|]|\Psi_{\text{ph}}\rangle-\langle\Psi_{\text{ph}}|\hat{W}|\Psi_{\text{ph}}\rangle.
\end{aligned}
\end{equation}
in which $\langle\Psi_{\text{ph}}|\hat{W}|\Psi_{\text{ph}}\rangle=\int \mathrm{d}\mathbf{x_{\text{p}}}\mathrm{d}\mathbf{x_{\text{p}}^{\prime}}\mathrm{d}\mathbf{x_{\text{h}}}\mathrm{d}\mathbf{x_{\text{h}}^{\prime}}\Psi_{\text{ph}}^{*}(\mathbf{x_{\text{p}}},\mathbf{x_{\text{h}}})W(\mathbf{x_{\text{h}}},\mathbf{x_{\text{h}}^{\prime}};\mathbf{x_{\text{p}}},\mathbf{x_{\text{p}}^{\prime}})\Psi_{\text{ph}}(\mathbf{x_{\text{p}}^{\prime}},\mathbf{x_{\text{h}}^{\prime}}).$
After extending the definition domain of $\Delta E[\Psi_{\text{ph}}(\mathbf{x}_{\text{p}},\mathbf{x}_{\text{h}})]$
to general particle-hole wavefunctions, a Hamiltonian for the particle-hole
wavefunction can be given as $\hat{H}|\Psi_{\text{ph}}\rangle=\frac{\delta\Delta E[\Psi_{\text{ph}}]}{\delta\langle\Psi_{\text{ph}}|}$.
In the basis of canonical particle-hole pairs, the Hamiltonian can
be written as

\begin{equation}
\begin{aligned}
&H_{(i\sigma,a\tau),(j\lambda,b\omega)}=\langle\Psi_{i\sigma,a\tau}|\hat{H}|\Psi_{j\lambda,b\omega}\rangle\\=&(\epsilon_{a\tau}^{\text{QP}}-\epsilon_{i\sigma}^{\text{QP}})\delta_{ij}\delta_{ab}\delta_{\sigma\lambda}\delta_{\tau\omega}-(\psi_{i\sigma}\psi_{j\lambda}|W|\psi_{b\omega}\psi_{a\tau}),\label{eq:unpurified_H}
\end{aligned}
\end{equation}
and the excitations energies can be obtained by diagonalizing the effective Hamiltonian,
$\sum_{j\lambda,b\omega}H_{(i\sigma,a\tau),(j\lambda,b\omega)}X_{j\lambda,b\omega}^{n}=\Delta E_{n}X_{i\sigma,a\tau}^{n}$.

This quasiparticle Hamiltonian (QH) inherits the broken triplet three-fold
degeneracy from $\Delta$SCF and OE, which can be solved by extending
the spin purification procedure in OE. To restore the triplet degeneracy,
the off-diagonal matrix element $H_{ia,\overline{jb}}$ ($H_{\overline{ia},jb}$)
is approximated as the difference between the single-determinant $S_{\text{z}}=0$
states and triplet states ($|\text{T},S_{\text{z}}=\pm1\rangle$),
i.e., $H_{ia,\overline{jb}}\equiv W_{i\overline{j},a\overline{b}}=W_{ij,ab}-W_{ij,\overline{ab}}$.
As a result, the spin structure of the quasiparticle Hamiltonian is
explicitly showed in equation \ref{eq:spin_adapted_ST}, when multiplied
by $(\langle\Psi_{ia}|,\langle\Psi_{i\overline{a}}|,\langle\Psi_{\overline{i}a}|,\langle\Psi_{\overline{ia}}|)$
on the left and $(|\Psi_{jb}\rangle,|\Psi_{j\overline{b}}\rangle,|\Psi_{\overline{j}b}\rangle,|\Psi_{\overline{jb}}\rangle)^{\text{T}}$
on the right,

\begin{equation}
\mathbf{H}=(\epsilon_{a}^{\text{QP}}-\epsilon_{i}^{\text{QP}})\delta_{ij}\delta_{ab}-\left(\begin{array}{cccc}
W_{ij,ab} & 0 & 0 & W_{i\overline{j},a\overline{b}}\\
0 & W_{ij,\overline{ab}} & 0 & 0\\
0 & 0 & W_{\overline{ij},ab} & 0\\
W_{\overline{i}j,\overline{a}b} & 0 & 0 & W_{\overline{ij},\overline{ab}}
\end{array}\right),\label{eq:spin_adapted_ST}
\end{equation}
the quasiparticle energy difference is a $4\times 4$ diagonal matrix.
An alternative way to restore the spin symmetry is to apply non-collinear
extension to the collinear exchange correlation kernels. Established
on perturbative derivatives of the parent DFA, the broken symmetry
of the quasiparticle Hamitonian in Eq. \ref{eq:unpurified_H} can
be attributed to intrinsic errors of the collinear DFAs. For collinear
DFAs, neither the energy functional nor the exchange correlation kernel
preserves the SU(2) symmetry of the exact Hamiltonian. The extension
to non-collinear kernels restores the SU(2) symmetry in spin-flip
(SF-) TDDFT formalism\citep{shao_spinflip_2003,wang_time-dependent_2004,wang_noncollinear_2025,wang_zero_2025}.
We show in supplementary materials\citep{SuppleMater} (section I.D) that including the non-collinear
exchange correlation kernels results in the same quasiparticle Hamiltonian
as the spin purification scheme just presented.

We now compare our QH with the BSE approach. Consider the $S_{\text{z}}=0$
block which couples singlet and triplet states. The eigenvalue equation
resembles BSE in the particle-hole channel, except for two key distinctions
on the quasiparticle energies $\epsilon_{a}^{\text{QP}}$, $\epsilon_{i}^{\text{QP}}$
and the quasiparticle interaction $W$. (1) The quasiparticle energies
in BSE are typically GW quasiparticle energies with dynamic screening
based on the Random Phase Approximation (RPA). The computational cost of
GW scales as $O(N^{4}-N^{6})$ w.r.t. system size\citep{hybertsen_electron_1986,ren_resolution--identity_2012,doi:10.1021/ct300648t,golze_core-level_2018,golze_gw_2019,panades-barrueta_accelerating_2023}.
In contrast, the quasiparticle energy in OE uses the static generalized
screened interaction, which has an essentially $O(N^{3})$ cost
(with one single quick O($N^{4}$) step), through Sherman-Morrison-based
Resolution of the Identity (RI) approximation\citep{fan_eliminating_2026}. In practice,
one-shot GW exhibits strong starting-point dependence\citep{bruneval_systematic_2015,jacquemin_benchmarking_2015},
while partially self-consistent GW mitigates at increased computational
cost\citep{blase_first-principles_2011}. The  screened
interaction leads to more consistent quasiparticle energies with
various DFA starting point. (2) Beyond the quasiparticle
energies, our quasiparticle Hamiltonian incorporates contributions from the exchange correlation kernel in the particle-hole interaction $W$, which may be related to the vertex corrections\citep{shishkin_accurate_2007}.
The exchange correlation kernel enters into the partial-hole interaction $W$ in two ways beyond the BSE screened Coulumb interaction: directly in $K$ and indirectly through the linear density matrix response function $\chi$ based on the DFA calculations.

Apart from these two distinctions, BSE approach predicts the singlet-triplet splitting to be determined by $2\left(ia|v|jb\right)$, which is known
to be too large\citep{bruneval_systematic_2015,blase_bethesalpeter_2018}.
In the ph-QH approach, the splitting is controlled by $2(W_{ij,ab}-W_{\overline{ij},ab})$,
consistent with spin purification in $\Delta$SCF. The form $2\left(ia|v|jb\right)$ can also be obtained in truncated CI methods such as CIS\citep{bene_selfconsistent_1971}.
The expression becomes exact in the full CI limit, whereas in the
truncated space the effective interaction might deviate from it. Our approach incorporates an effective interaction through the generalized screened interaction $W$. For the effective interaction in the CIS space, contributions from double and higher excitations might be mapped into dynamically screening interaction, similar to non-adiabatic TDDFT\citep{ferre_density-functional_2016}.

A technical consideration in the determination of GSC2 quasiparticle
energy is that $f_{\text{xc}}$ integrals may become numerically unstable
for virtual states with significant diffuse character\citep{zheng_improving_2011,li_piecewise_2017}.
For example, the LDA exchange functional $E_{\text{x}}^{\text{LDA}}\sim-\int\mathrm{d}\mathbf{r}\rho^{\frac{4}{3}}(\mathbf{r})$
produces an exchange kernel proportional to $\rho^{-\frac{2}{3}}(\mathbf{r})\delta(\mathbf{r}-\mathbf{r}^{\prime})$.
For sufficiently diffuse virtual orbitals p, q, the exchange-correlation integral, $\left(pp|f_{\text{xc}}^{\alpha\alpha}|qq\right)$, can numerically diverge. This may be related to some deficiency of local or semi-local functionals. In the previous OE work, this issue is mitigated by evaluating the curvature matrix $\kappa$ at a small finite fractional
occupation away from the integer point, however with the fixed orbitals
from the integer calculations for computational efficiency\citep{fan_OE_2026}. This strategy solves the negative curvature problem for OE,
since only the quasiparticle energies of the low-lying states are needed. In this work, since a full spectrum of quasiparticle energies
are required, this strategy becomes less practical computationally.
To address this issue, the pure functional part of the exchange-correlation
kernel is removed from $K$, yielding $K_{pq,rs}^{\prime}=(pq|sr)-c^{\text{HF}}_{\text{x}}(pr|sq)$.
Therefore the quasiparticle energy of the virtual states are expressed as shown below,

\begin{equation}
\begin{aligned}
\kappa_{a\sigma,a\sigma}\approx &K_{aa\sigma,aa\sigma}^{\text{\ensuremath{\prime}}}+\sum_{ib\tau,jc\lambda}[(K_{aa\sigma,ib\tau}^{\text{\ensuremath{\prime}}}+K_{aa\sigma,bi\tau}^{\text{\ensuremath{\prime}}})\chi_{ib\tau,jc\lambda}\\ &K_{jc\lambda,aa\sigma}^{\text{\ensuremath{\prime}}}],\quad a\sigma\in\text{virt.} \label{eq:regularized}
\end{aligned}
\end{equation}

Since no ill-defined integral is involved in $\chi$ in Eq. \ref{eq:regularized}, the density (matrix) linear response function $\chi$ remains unchanged. Due to the existence
of occupied orbitals in the integration, which makes the integrals well behaved, no regularization is applied to $W_{ij,ab}$. The treatment is numerically more stable, while preserving
the overall accuracy of quasiparticle energy. The ph-QH approach based
on regularized generalized screened interaction for the quasiparticle energies of virtual states is referred to as ph-QH@DFA, since $\kappa_{p\sigma,p\sigma}$ directly originates from the DFA second-order derivatives. The effectiveness of this approximation can be assessed by quasiparticle energies. Table \ref{table:HOMO_LUMO_gap} reports the mean signed errors (MSEs) and mean absolute errors (MAEs) of the HOMO-LUMO quasiparticle energy gap obtained with evGW, and GSC2, with respect to CCSD(T)/TZVP reference. For these systems, the LUMOs are typically unbounded states, therefore the theoretical values are used as the reference.

\begin{table}
\centering \resizebox{0.47\textwidth}{!}{%
\begin{tabular}{ccccccc}
\toprule 
 & \multicolumn{3}{c}{PBE} & \multicolumn{3}{c}{B3LYP}\tabularnewline
\cmidrule(lr){2-4}\cmidrule(lr){5-7}
 & GSC2 & lrLOSC & evGW & GSC2 & lrLOSC & evGW\tabularnewline
\midrule 
MSE & -0.22 & 0.16 & -0.03 & -0.20 & -0.01 & -0.05\tabularnewline
MAE & 0.39 & 0.29 & 0.19 & 0.27 & 0.23 & 0.16\tabularnewline
\bottomrule
\end{tabular}} \caption{ The MSEs and MAEs of HOMO-LUMO quasiparticle energy gaps for (determined
by GSC2, lrLOSC and evGW method, for small organic molecules\citep{schreiber_benchmarks_2008,silva-junior_benchmarks_2010}.
The GSC2 quasiparticle energy is equivalent to the OE theory when
applied to the single-particle channel or single-hole channel. The column headers are organized in two hierarchical levels. The upper level specifies the DFA starting point (PBE or B3LYP), while the lower level indicates the quasiparticle scheme (GSC2, lrLOSC, or evGW). The basis adopted
is TZVP, and the reference is the CCSD(T)/TZVP calculation\citep{bruneval_systematic_2015}.
All numbers are in the unit of eV.\label{table:HOMO_LUMO_gap}}
\end{table}

\begin{figure}
\includegraphics[width=0.49\linewidth]{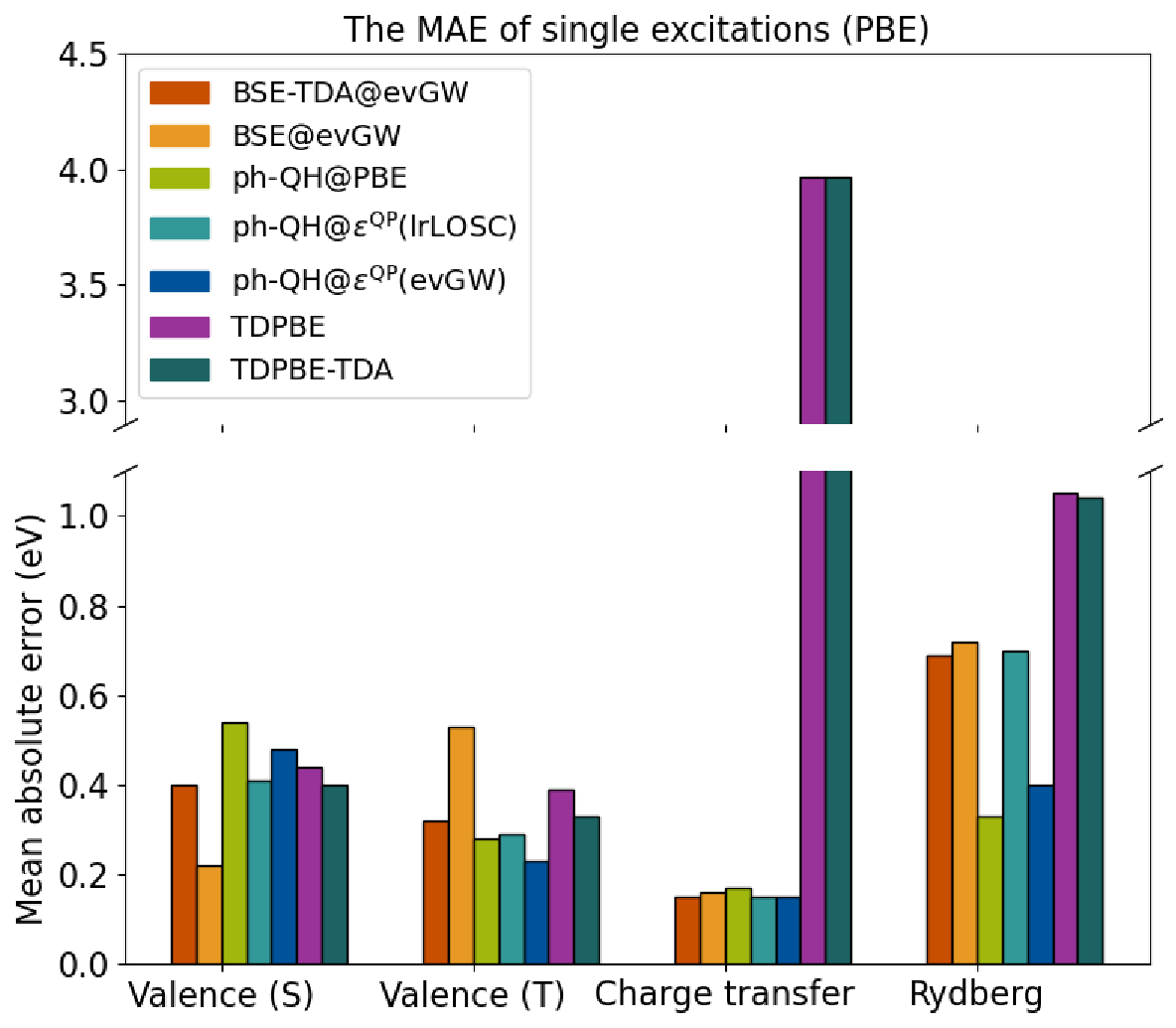}
\includegraphics[width=0.49\linewidth]{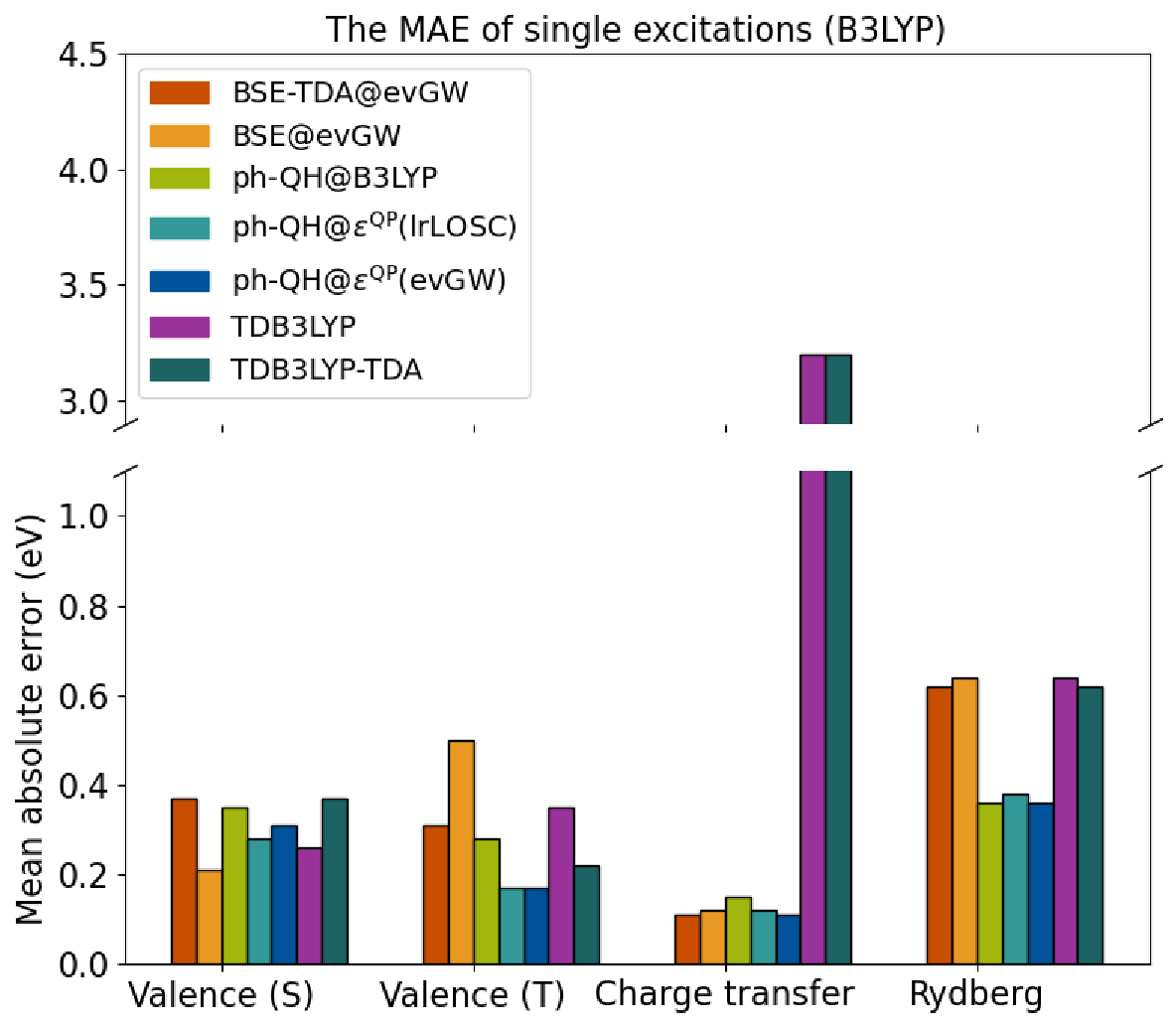}\caption{The MAEs of the valence singlet (S) excitations, valence triplet (T)
excitations, CT and Rydberg excitations. All calculation utilizes
the PBE\citep{perdew_generalized_1996} or B3LYP\citep{lee_development_1988}
functional for the parent KS-DFT calculation. For the valence singlet
and triplet excitations, the tests are performed on small organic
molecules\citep{schreiber_benchmarks_2008,silva-junior_benchmarks_2010},
with the TZVP basis set\citep{schafer_fully_1992}. For charge-transfer
excitations, the tested systems include 13 intermolecular excitations
\citep{kozma_new_2020} with theoretical best estimates and 4 excitations
of the gas-phase molecular complex with experimental values\citep{stein_reliable_2009}.
A double-zeta basis set cc-pVDZ\citep{dunning_gaussian_1989} is utilized
to avoid the interference of Rydberg states. For the Rydberg excitations,
the examination is performed on small organic molecules and atoms\citep{loos_mountaineering_2018},
with Dunning's augmented correlation consistent triple-zeta basis
set aug-cc-pVTZ\citep{dunning_gaussian_1989,kendall_electron_1992}.
See supporting information for basis set convergence of Rydberg excitations\citep{SuppleMater}.}\label{fig:MAEs_single}
\end{figure}

To assess the accuracy of the particle-hole quasiparticle Hamiltonian, we perform
numerical validation across a broad set of neutral molecular excitations,
including the valence, CT and Rydberg excitations. The statistical
results are shown in Fig. \ref{fig:MAEs_single}. Regardless of the DFA starting point, ph-QH@DFA performs well similarly  to BSE(-TDA)@evGW on charge transfer excitations, and outperforms BSE(-TDA)@evGW on Rydberg excitations.

For the valence singlet excitations, the ph-QH@PBE results are worse
than BSE-TDA by 0.15 eV. The ph-QH shows a MSE of -0.44 eV, while
BSE-TDA manifests a MSE of 0.40 eV. While this error remaining acceptable,
the underestimation is related to the DFA delocalization error\citep{perdew_self-interaction_1981,mori-sanchez_localization_2008},
and can be further improved by adopting localized orbital scaling
correction (LOSC) for better quasiparticle energies than those given
by OE or GSC2 \citep{li_localized_2018,su_preserving_2020,mei_self-consistent_2020,mei_exact_2021,yu_accurate_2025}.
With the regularization scheme applied, the linear-response lrLOSC@PBE
quasiparticle energy manifests a MSE of 0.16 eV and a MAE of 0.23
eV, as shown in Table \ref{table:HOMO_LUMO_gap}. With lrLOSC quasiparticle
energy, the ph-QH metod, denoted as ph-QH@$\epsilon^{\text{QP}}(\text{lrLOSC})$,
yields a MAE of 0.40 eV and 0.35 eV for valence singlet and triplet
excitations. For B3LYP starting point, regularized lrLOSC quasiparticle
energy shows -0.01 eV MSE and 0.23 eV MAE. The accurate regularized
lrLOSC@B3LYP quasiparticle energies lead to a MAE of 0.28 eV in the
valence singlet excitations, and a MAE of 0.17 eV in the valence triplet
excitations. Detailed information can be found in the supplementary
materials \citep{SuppleMater} (Table II and Table III).

To facilitate a more direct comparison to BSE@evGW, ph-QH@$\epsilon^{\text{QP}}(\text{evGW})$ (only replacing the quasiparticle energy with evGW quasiparticle energy in Eq. \ref{eq:spin_adapted_ST}) results are also reported in Fig. \ref{fig:MAEs_single}.
For all tested functionals and each tested type of single excitations,
the ph-QH@evGW results are at least comparable to BSE(-TDA)@evGW. The better performance than BSE(-TDA) on Rydberg and valence triplet excitations is also observed for ph-QH@evGW method.
The comparison of ph-QH@DFA, ph-QH@$\epsilon^{\text{QP}}(\text{lrLOSC})$
and ph-QH@$\epsilon^{\text{QP}}(\text{evGW})$ to BSE(-TDA)@evGW establishes
the validity of the derived quasiparticle Hamiltonian. More details
on each individual excitation, can be found in the supporting information \citep{SuppleMater}
(section III.A-C).

\begin{figure}
\centering \includegraphics[width=0.95\linewidth]{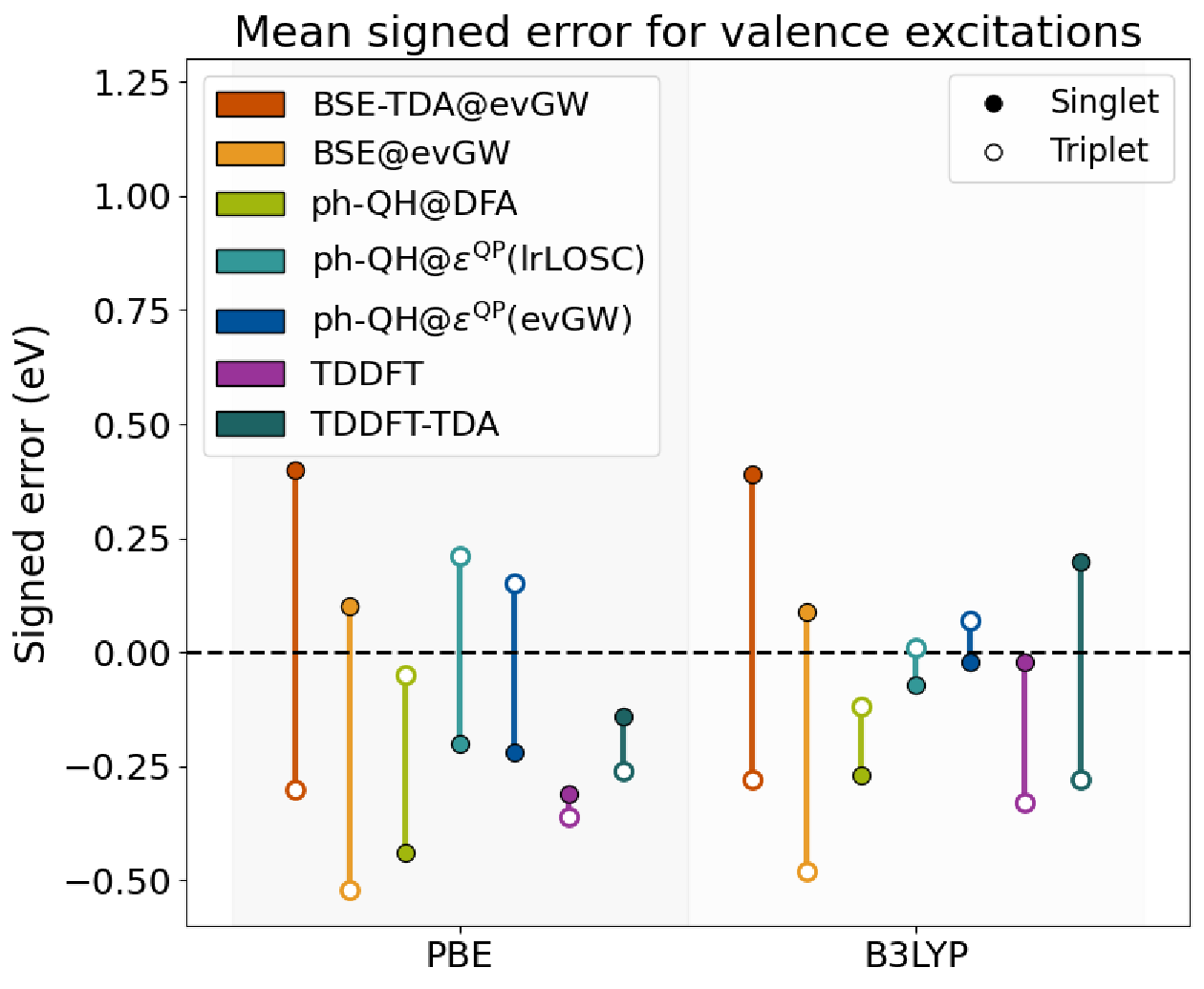}
\caption{The mean signed errors of singlet and triplet excitations for small
organic molecules \citep{schreiber_benchmarks_2008,silva-junior_benchmarks_2010},
with PBE/B3LYP DFA starting point\citep{lee_development_1988,perdew_generalized_1996}.
The full circles represent the singlet errors, and the open circles
are for the triplet errors. The ph-BSE calculation performed on top
of evGW quasiparticle energy is labelled as ``BSE@evGW''. The BSE
with TDA approximation is labelled as ``BSE-TDA''. TDDFT calculations
with and without TDA (``TDDFT'' and ``TDDFT-TDA'') are also presented
for comparison. As indicated by the solid line connecting singlet
error and triplet error, BSE/BSE-TDA overestimates the splitting between
singlets and triplets. The basis set adopted is TZVP.}\label{fig:ST_split}
\end{figure}

As discussed above, the singlet-triplet splitting is typically overestimated
by BSE approach, which is recognized as an intrinsic error in BSE\citep{bruneval_systematic_2015,blase_bethesalpeter_2018},
even with accurate quasiparticle energies. In this work, the MSEs of
valence singlet and triplet excitations are analyzed in Fig. \ref{fig:ST_split}.
Compared to BSE-TDA@evGW, the ph-QH approach has a better estimate
of triplet energies, while maintaining the quality of singlet excitations.
As shown in Fig \ref{fig:ST_split}, the particle-hole Hamiltonian approach gives
a greatly enhanced prediction on the difference between singlet MSE
and triplet MSE, -0.15 eV, for the B3LYP starting point. For the PBE
starting point, the splitting is underestimated by 0.30-0.40
eV, yet this remains substantially more accurate than the 0.70 eV
overestimation from BSE(-TDA)@evGW@PBE or BSE(-TDA)@evGW@B3LYP.

As described above, $\Delta$SCF and OE methods are not suitable for
describing multi-configurational excited states, since these methods
are limited to single-determinant excited-state wavefunction. For
excited-state wavefunction with symmetry-protected multi-configurational
nature, OE typically leads to qualitatively incorrect predictions.
For instance, the $\pi_{E_{1g}}$ to $\pi_{E_{2u}}^{*}$ transition
of the benzene molecule should be decomposed as $E_{1g}\otimes E_{2u}=B_{1u}\oplus B_{2u}\oplus E_{1u}$.
This decomposition indicates an excitation energy with 2-fold degeneracy
and two separate states. However, since all of the excited states
($B_{1u}$, $B_{2u}$, $E_{1u}$) are essentially multi-configurational,
none of these wave functions are reproduced by OE or $\Delta$SCF.
OE predicts two 2-fold degeneracy. With the quasiparticle Hamiltonian
extension of OE, the results are quantitatively in good agreement
with theoretical best estimates. See supporting information\citep{SuppleMater} (Table XXX) for more
details on excitation energies and a complete set of degeneracy test.

Another category of intrinsically multi-configurational excited states
are the plasmon-like excitations\citep{bernadotte_plasmons_2013,krauter_plasmons_2014,krauter_identification_2015,langford_plasmon_2021}.
In conjugated polyenes (butadiene, hexatriene and octatetraene), the
valence orbitals manifest plane-wave-like feature. Transitions between
these valence orbitals with the same momentum change are of the same
symmetry, and may linearly combine to yield the excited-state wave
functions. For the genuine plasmon excitations, these CIS determinants
combine in a constructive way\citep{langford_plasmon_2021}. For the
remaining excitations, the multi-configurational feature pertains,
and they might be identified as plasmon-like excitations. These intrinsically
multi-configurational states, cannot be captured by $\Delta$SCF or
OE with great accuracy. For instance, for $1^{3}B_{u}$ and $1^{1}B_{u}$ transition of butadiene dominated by the HOMO-LUMO transition, the OE excitation energy is overestimated by 0.56 eV and 0.03 eV, respectively. Nevertheless,
for the $1^{3}A_{g}$ transition of butadiene, as a linear combination of HOMO-1 to LUMO and HOMO to LUMO+1 transition, the OE excitation energy is severely overestimated by 1.45 eV. The 1$^{3}A_{g}$ excitation of hexatriene and octatetraene similarly shows more errors than the 1$^{3}B_{u}$, 1$^{1}B_{u}$ states.
These drawbacks are mitigated by the particle-hole quasiparticle Hamiltonian approach presented in this work. As shown in the supplementary materials\citep{SuppleMater} (Table XXIX), for the ph-QH approach, the errors for these 9 excitations are all within 0.3 eV.

The oscillator strength associated with the singlet excitation for
small organic molecules is examined as a manifestation of the quality
of excited-state wavefunction. The comparison of the oscillator strength
determined by this particle-hole Hamiltonian to the BSE and TDDFT,
w.r.t. CASPT2 reference and CC3 reference, can be found in the supporting
information\citep{SuppleMater} (section III.E). The results are qualitatively in good agreement to the
BSE-TDA results.

To summarize, in this work, we develop the OE excitation energy function
into a quasiparticle Hamiltonian. For particle-hole excitations, the performance
on both the excitation energy and the associated oscillator strength
is examined. For all types of excitations, this quasiparticle Hamiltonian
yields results similar to or better than BSE. This method shares a
similar description of particle-hole interaction with BSE, but with
additional consideration of $f_{\text{xc}}$ both in the interaction
and in its screening with the density linear response. By utilizing
a matrix form of spin purification, this method gives a better estimation
of singlet-triplet splitting than BSE. Furthermore, the statistical
error for the Rydberg excitations is notably improved compared to
BSE. In contrast to OE and $\Delta$SCF, this approach is capable of capturing excited
state with multi-configurational feature, which should make it applicable
to general optical excitations in bulk materials, including excitonic
states. It gives a correct prediction of the splitting mode for symmetry-protected
excitations, and yielding better estimations of plasmon-like excitations.
The quasiparticle Hamiltonian generated from OE can include multiple
particles and multiple holes as well, with the potential applicability
to double and multiple excitations.

Details of quasiparticle derivation, validation of regularization treatment, computational details, and the complete benchmark data for all systems studied are included in the supplementary material\citep{SuppleMater} (See also references \citep{blum_ab_2009, cohen_fractional_spin_2008, jacquemin_assessment_2016, li_localized_2025, martin_natural_2003, Momma:db5098, PhysRevLett.49.1691, PhysRevA.76.040501, su_describing_2018, https://doi.org/10.1002/jcc.23981,sun_pyscf_2018, sun_recent_2020, 10.1063/5.0285025, PhysRevLett.84.5172, zhang_perspective_2000}) therein.

We acknowledge support from the National Institute of Health (1R35GM158181-01).

\bibliography{GW-BSE_new}

@article{golze_gw_2019,
	title = {The {GW} {Compendium}: {A} {Practical} {Guide} to {Theoretical} {Photoemission} {Spectroscopy}},
	volume = {7},
	shorttitle = {The {GW} {Compendium}},
	urldate = {2024-03-22},
	journal = {Front. Chem.},
	author = {Golze, Dorothea and Dvorak, Marc and Rinke, Patrick},
	year = {2019},
}

@article{ren_resolution--identity_2012,
	title = {Resolution-of-identity approach to {Hartree}–{Fock}, hybrid density functionals, {RPA}, {MP2} and {GW} with numeric atom-centered orbital basis functions},
	volume = {14},
	number = {5},
	urldate = {2024-03-11},
	journal = {New J. Phys.},
	author = {Ren, Xinguo and Rinke, Patrick and Blum, Volker and Wieferink, Jürgen and Tkatchenko, Alexandre and Sanfilippo, Andrea and Reuter, Karsten and Scheffler, Matthias},
	year = {2012},
	pages = {053020},
}

@article{golze_core-level_2018,
	title = {Core-{Level} {Binding} {Energies} from {GW}: {An} {Efficient} {Full}-{Frequency} {Approach} within a {Localized} {Basis}},
	volume = {14},
	shorttitle = {Core-{Level} {Binding} {Energies} from {GW}},
	number = {9},
	urldate = {2025-04-29},
	journal = {J. Chem. Theory Comput.},
	author = {Golze, Dorothea and Wilhelm, Jan and van Setten, Michiel J. and Rinke, Patrick},
	year = {2018},
	pages = {4856--4869},
}

@article{kozma_new_2020,
	title = {A {New} {Benchmark} {Set} for {Excitation} {Energy} of {Charge} {Transfer} {States}: {Systematic} {Investigation} of {Coupled} {Cluster} {Type} {Methods}},
	volume = {16},
	shorttitle = {A {New} {Benchmark} {Set} for {Excitation} {Energy} of {Charge} {Transfer} {States}},
	number = {7},
	urldate = {2025-04-30},
	journal = {J. Chem. Theory Comput.},
	author = {Kozma, Balázs and Tajti, Attila and Demoulin, Baptiste and Izsák, Róbert and Nooijen, Marcel and Szalay, Péter G.},
	year = {2020},
	pages = {4213--4225},
}

@article{loos_mountaineering_2018,
	title = {A {Mountaineering} {Strategy} to {Excited} {States}: {Highly} {Accurate} {Reference} {Energies} and {Benchmarks}},
	volume = {14},
	shorttitle = {A {Mountaineering} {Strategy} to {Excited} {States}},
	number = {8},
	urldate = {2025-04-30},
	journal = {J. Chem. Theory Comput.},
	author = {Loos, Pierre-François and Scemama, Anthony and Blondel, Aymeric and Garniron, Yann and Caffarel, Michel and Jacquemin, Denis},
	year = {2018},
	pages = {4360--4379},
}

@article{krause_implementation_2017,
	title = {Implementation of the {Bethe}-{Salpeter} equation in the {TURBOMOLE} program},
	volume = {38},
	number = {6},
	urldate = {2025-05-01},
	journal = {J. Comput. Chem.},
	author = {Krause, Katharina and Klopper, Wim},
	year = {2017},
	pages = {383--388},
}

@article{silva-junior_benchmarks_2010,
	title = {Benchmarks of electronically excited states: {Basis} set effects on {CASPT2} results},
	volume = {133},
	shorttitle = {Benchmarks of electronically excited states},
	number = {17},
	urldate = {2025-05-02},
	journal = {J. Chem. Phys.},
	author = {Silva-Junior, Mario R. and Schreiber, Marko and Sauer, Stephan P. A. and Thiel, Walter},
	year = {2010},
	pages = {174318},
}

@article{schreiber_benchmarks_2008,
	title = {Benchmarks for electronically excited states: {CASPT2}, {CC2}, {CCSD}, and {CC3}},
	volume = {128},
	shorttitle = {Benchmarks for electronically excited states},
	number = {13},
	urldate = {2025-05-02},
	journal = {J. Chem. Phys.},
	author = {Schreiber, Marko and Silva-Junior, Mario R. and Sauer, Stephan P. A. and Thiel, Walter},
	year = {2008},
	pages = {134110},
}

@article{blase_bethesalpeter_2018,
	title = {The {Bethe}–{Salpeter} equation in chemistry: relations with {TD}-{DFT}, applications and challenges},
	volume = {47},
	shorttitle = {The {Bethe}–{Salpeter} equation in chemistry},
	number = {3},
	urldate = {2025-05-06},
	journal = {Chem. Soc. Rev.},
	author = {Blase, Xavier and Duchemin, Ivan and Jacquemin, Denis},
	year = {2018},
	pages = {1022--1043},
}

@article{jacquemin_benchmarking_2015,
	title = {Benchmarking the {Bethe}–{Salpeter} {Formalism} on a {Standard} {Organic} {Molecular} {Set}},
	volume = {11},
	number = {7},
	urldate = {2025-05-06},
	journal = {J. Chem. Theory Comput.},
	author = {Jacquemin, Denis and Duchemin, Ivan and Blase, Xavier},
	year = {2015},
	pages = {3290--3304},
}

@article{shishkin_accurate_2007,
	title = {Accurate {Quasiparticle} {Spectra} from {Self}-{Consistent} {GW} {Calculations} with {Vertex} {Corrections}},
	volume = {99},
	number = {24},
	urldate = {2025-05-08},
	journal = {Phys. Rev. Lett.},
	author = {Shishkin, M. and Marsman, M. and Kresse, G.},
	year = {2007},
	pages = {246403},
}

@article{bruneval_systematic_2015,
	title = {A systematic benchmark of the ab initio {Bethe}-{Salpeter} equation approach for low-lying optical excitations of small organic molecules},
	volume = {142},
	number = {24},
	urldate = {2025-05-12},
	journal = {J. Chem. Phys.},
	author = {Bruneval, Fabien and Hamed, Samia M. and Neaton, Jeffrey B.},
	year = {2015},
	pages = {244101},
}

@article{jacquemin_assessment_2016,
	title = {Assessment of the {Accuracy} of the {Bethe}–{Salpeter} ({BSE}/{GW}) {Oscillator} {Strengths}},
	volume = {12},
	number = {8},
	urldate = {2025-05-12},
	journal = {J. Chem. Theory Comput.},
	author = {Jacquemin, Denis and Duchemin, Ivan and Blondel, Aymeric and Blase, Xavier},
	year = {2016},
	pages = {3969--3981},
}

@article{van_setten_gw100_2015,
	title = {{GW100}: {Benchmarking} {G0W0} for {Molecular} {Systems}},
	volume = {11},
	shorttitle = {{GW100}},
	number = {12},
	urldate = {2025-05-14},
	journal = {J. Chem. Theory Comput.},
	author = {van Setten, Michiel J. and Caruso, Fabio and Sharifzadeh, Sahar and Ren, Xinguo and Scheffler, Matthias and Liu, Fang and Lischner, Johannes and Lin, Lin and Deslippe, Jack R. and Louie, Steven G. and Yang, Chao and Weigend, Florian and Neaton, Jeffrey B. and Evers, Ferdinand and Rinke, Patrick},
	year = {2015},
	pages = {5665--5687},
}

@article{jacquemin_is_2017,
	title = {Is the {Bethe}–{Salpeter} {Formalism} {Accurate} for {Excitation} {Energies}? {Comparisons} with {TD}-{DFT}, {CASPT2}, and {EOM}-{CCSD}},
	volume = {8},
	shorttitle = {Is the {Bethe}–{Salpeter} {Formalism} {Accurate} for {Excitation} {Energies}?},
	number = {7},
	urldate = {2025-05-14},
	journal = {J. Phys. Chem. Lett.},
	author = {Jacquemin, Denis and Duchemin, Ivan and Blase, Xavier},
	year = {2017},
	pages = {1524--1529},
}

@article{mori-sanchez_localization_2008,
	title = {Localization and {Delocalization} {Errors} in {Density} {Functional} {Theory} and {Implications} for {Band}-{Gap} {Prediction}},
	volume = {100},
	number = {14},
	urldate = {2025-05-26},
	journal = {Phys. Rev. Lett.},
	author = {Mori-Sánchez, Paula and Cohen, Aron J. and Yang, Weitao},
	year = {2008},
	pages = {146401},
}

@article{mei_self-consistent_2020,
	title = {Self-{Consistent} {Calculation} of the {Localized} {Orbital} {Scaling} {Correction} for {Correct} {Electron} {Densities} and {Energy}-{Level} {Alignments} in {Density} {Functional} {Theory}},
	volume = {11},
	number = {23},
	urldate = {2025-05-26},
	journal = {J. Phys. Chem. Lett.},
	author = {Mei, Yuncai and Chen, Zehua and Yang, Weitao},
	year = {2020},
	pages = {10269--10277},
}

@article{zheng_improving_2011,
	title = {Improving {Band} {Gap} {Prediction} in {Density} {Functional} {Theory} from {Molecules} to {Solids}},
	volume = {107},
	number = {2},
	urldate = {2025-06-08},
	journal = {Phys. Rev. Lett.},
	author = {Zheng, Xiao and Cohen, Aron J. and Mori-Sánchez, Paula and Hu, Xiangqian and Yang, Weitao},
	year = {2011},
	pages = {026403},
}

@article{loos_dynamical_2020,
	title = {Dynamical correction to the {Bethe}–{Salpeter} equation beyond the plasmon-pole approximation},
	volume = {153},
	number = {11},
	urldate = {2025-07-08},
	journal = {J. Chem. Phys.},
	author = {Loos, Pierre-François and Blase, Xavier},
	year = {2020},
	pages = {114120},
}

@article{martin_natural_2003,
	title = {Natural transition orbitals},
	volume = {118},
	number = {11},
	urldate = {2025-08-04},
	journal = {J. Chem. Phys.},
	author = {Martin, Richard L.},
	year = {2003},
	pages = {4775--4777},
}

@article{langford_plasmon_2021,
	title = {Plasmon {Character} {Index}: {An} {Accurate} and {Efficient} {Metric} for {Identifying} and {Quantifying} {Plasmons} in {Molecules}},
	volume = {12},
	shorttitle = {Plasmon {Character} {Index}},
	number = {38},
	urldate = {2025-08-13},
	journal = {J. Phys. Chem. Lett.},
	author = {Langford, James and Xu, Xi and Yang, Yang},
	year = {2021},
	pages = {9391--9397},
}

@article{10.1063/5.0260895,
    author = {Hillenbrand, Christopher and Li, Jiachen and Zhu, Tianyu},
    title = {Energy-specific Bethe–Salpeter equation implementation for efficient optical spectrum calculations},
    journal = {J. Chem. Phys.},
    volume = {162},
    number = {17},
    pages = {174117},
    year = {2025},
}

@article{doi:10.1021/ct300648t,
author = {van Setten, M. J. and Weigend, F. and Evers, F.},
title = {The GW-Method for Quantum Chemistry Applications: Theory and Implementation},
journal = {J. Chem. Theory Comput.},
volume = {9},
number = {1},
pages = {232-246},
year = {2013},
}

@article{krauter_identification_2015,
	title = {Identification of {Plasmons} in {Molecules} with {Scaled} {Ab} {Initio} {Approaches}},
	volume = {119},
	number = {43},
	urldate = {2025-06-20},
	journal = {J. Phys. Chem. C},
	author = {Krauter, Caroline M. and Bernadotte, Stephan and Jacob, Christoph R. and Pernpointner, Markus and Dreuw, Andreas},
	year = {2015},
	pages = {24564--24573},
}

@article{krauter_plasmons_2014,
	title = {Plasmons in molecules: {Microscopic} characterization based on orbital transitions and momentum conservation},
	volume = {141},
	shorttitle = {Plasmons in molecules},
	number = {10},
	urldate = {2025-06-20},
	journal = {J. Chem. Phys.},
	author = {Krauter, Caroline M. and Schirmer, Jochen and Jacob, Christoph R. and Pernpointner, Markus and Dreuw, Andreas},
	year = {2014},
	pages = {104101},
}

@article{bernadotte_plasmons_2013,
	title = {Plasmons in {Molecules}},
	volume = {117},
	number = {4},
	urldate = {2024-06-21},
	journal = {J. Phys. Chem. C},
	author = {Bernadotte, Stephan and Evers, Ferdinand and Jacob, Christoph R.},
	year = {2013},
	pages = {1863--1878},
}

@article{stein_reliable_2009,
	title = {Reliable {Prediction} of {Charge} {Transfer} {Excitations} in {Molecular} {Complexes} {Using} {Time}-{Dependent} {Density} {Functional} {Theory}},
	volume = {131},
	number = {8},
	urldate = {2025-06-17},
	journal = {J. Am. Chem. Soc.},
	author = {Stein, Tamar and Kronik, Leeor and Baer, Roi},
	year = {2009},
	pages = {2818--2820},
}

@article{wang_zero_2025,
	title = {Zero {Excitation} {Energy} {Theorem} and the {Spin}-{Flip} {Kernel}},
	volume = {21},
	number = {14},
	urldate = {2025-08-04},
	journal = {J. Chem. Theory Comput.},
	author = {Wang, Tai and Li, Hao and Gao, Yi Qin and Xiao, Yunlong},
	year = {2025},
	pages = {6905--6921},
}

@article{wang_noncollinear_2025,
	title = {Noncollinear generalization of nonlocal pure exchange–correlation functionals},
	volume = {162},
	number = {21},
	urldate = {2025-08-04},
	journal = {J. Chem. Phys.},
	author = {Wang, Tai and Li, Hao and Pu, Zhichen and Gao, Yi Qin and Xiao, Yunlong},
	year = {2025},
	pages = {214104},
}

@article{dreuw_single-reference_2005,
	title = {Single-{Reference} ab {Initio} {Methods} for the {Calculation} of {Excited} {States} of {Large} {Molecules}},
	volume = {105},
	number = {11},
	urldate = {2024-05-24},
	journal = {Chem. Rev.},
	author = {Dreuw, Andreas and Head-Gordon, Martin},
	year = {2005},
	pages = {4009--4037},
}

@article{dreuw_failure_2004,
	title = {Failure of {Time}-{Dependent} {Density} {Functional} {Theory} for {Long}-{Range} {Charge}-{Transfer} {Excited} {States}: {The} {Zincbacteriochlorin}-{Bacteriochlorin} and {Bacteriochlorophyll}-{Spheroidene} {Complexes}},
	volume = {126},
	shorttitle = {Failure of {Time}-{Dependent} {Density} {Functional} {Theory} for {Long}-{Range} {Charge}-{Transfer} {Excited} {States}},
	number = {12},
	urldate = {2025-03-20},
	journal = {J. Am. Chem. Soc.},
	author = {Dreuw, Andreas and Head-Gordon, Martin},
	year = {2004},
	pages = {4007--4016},
}

@article{lischner_first-principles_2012,
	title = {First-{Principles} {Calculations} of {Quasiparticle} {Excitations} of {Open}-{Shell} {Condensed} {Matter} {Systems}},
	volume = {109},
	copyright = {http://link.aps.org/licenses/aps-default-license},
	number = {3},
	urldate = {2024-05-08},
	journal = {Phys. Rev. Lett.},
	author = {Lischner, Johannes and Deslippe, Jack and Jain, Manish and Louie, Steven G.},
	year = {2012},
	pages = {036406},
}

@article{shao_spinflip_2003,
	title = {The spin–flip approach within time-dependent density functional theory: {Theory} and applications to diradicals},
	volume = {118},
	shorttitle = {The spin–flip approach within time-dependent density functional theory},
	number = {11},
	urldate = {2025-03-04},
	journal = {J. Chem. Phys.},
	author = {Shao, Yihan and Head-Gordon, Martin and Krylov, Anna I.},
	year = {2003},
	pages = {4807--4818},
}

@article{besley_self-consistent-field_2009,
	title = {Self-consistent-field calculations of core excited states},
	volume = {130},
	number = {12},
	urldate = {2025-03-20},
	journal = {J. Chem. Phys.},
	author = {Besley, Nicholas A. and Gilbert, Andrew T. B. and Gill, Peter M. W.},
	year = {2009},
	pages = {124308},
}

@article{hu_accelerating_2020,
	title = {Accelerating {Excitation} {Energy} {Computation} in {Molecules} and {Solids} within {Linear}-{Response} {Time}-{Dependent} {Density} {Functional} {Theory} via {Interpolative} {Separable} {Density} {Fitting} {Decomposition}},
	volume = {16},
	number = {2},
	urldate = {2025-03-20},
	journal = {J. Chem. Theory Comput.},
	author = {Hu, Wei and Liu, Jie and Li, Yingzhou and Ding, Zijing and Yang, Chao and Yang, Jinlong},
	year = {2020},
	pages = {964--973},
}

@article{gilbert_self-consistent_2008,
	title = {Self-{Consistent} {Field} {Calculations} of {Excited} {States} {Using} the {Maximum} {Overlap} {Method} ({MOM})},
	volume = {112},
	number = {50},
	urldate = {2025-03-28},
	journal = {J. Phys. Chem. A},
	author = {Gilbert, Andrew T. B. and Besley, Nicholas A. and Gill, Peter M. W.},
	year = {2008},
	pages = {13164--13171},
}

@article{stanton_equation_1993,
	title = {The equation of motion coupled‐cluster method. {A} systematic biorthogonal approach to molecular excitation energies, transition probabilities, and excited state properties},
	volume = {98},
	number = {9},
	urldate = {2025-03-30},
	journal = {J. Chem. Phys.},
	author = {Stanton, John F. and Bartlett, Rodney J.},
	year = {1993},
	pages = {7029--7039},
}

@misc{yang_foundation_2024,
	title = {Foundation for the {$\Delta$SCF} {Approach} in {Density} {Functional} {Theory}},
	urldate = {2025-04-06},
	author = {Yang, Weitao and Ayers, Paul W.},
	year = {2024},
	note = {arXiv:2403.04604 [physics]},
}

@article{sun_pyscf_2018,
	title = {{PySCF}: the {Python}-based simulations of chemistry framework},
	volume = {8},
	copyright = {© 2017 Wiley Periodicals, Inc.},
	shorttitle = {{PySCF}},
	number = {1},
	urldate = {2025-04-07},
	journal = {WIREs Comput. Mol. Sci.},
	author = {Sun, Qiming and Berkelbach, Timothy C. and Blunt, Nick S. and Booth, George H. and Guo, Sheng and Li, Zhendong and Liu, Junzi and McClain, James D. and Sayfutyarova, Elvira R. and Sharma, Sandeep and Wouters, Sebastian and Chan, Garnet Kin-Lic},
	year = {2018},
	pages = {e1340},
}

@article{sun_recent_2020,
	title = {Recent developments in the {PySCF} program package},
	volume = {153},
	number = {2},
	urldate = {2025-04-07},
	journal = {J. Chem. Phys.},
	author = {Sun, Qiming and Zhang, Xing and Banerjee, Samragni and Bao, Peng and Barbry, Marc and Blunt, Nick S. and Bogdanov, Nikolay A. and Booth, George H. and Chen, Jia and Cui, Zhi-Hao and Eriksen, Janus J. and Gao, Yang and Guo, Sheng and Hermann, Jan and Hermes, Matthew R. and Koh, Kevin and Koval, Peter and Lehtola, Susi and Li, Zhendong and Liu, Junzi and Mardirossian, Narbe and McClain, James D. and Motta, Mario and Mussard, Bastien and Pham, Hung Q. and Pulkin, Artem and Purwanto, Wirawan and Robinson, Paul J. and Ronca, Enrico and Sayfutyarova, Elvira R. and Scheurer, Maximilian and Schurkus, Henry F. and Smith, James E. T. and Sun, Chong and Sun, Shi-Ning and Upadhyay, Shiv and Wagner, Lucas K. and Wang, Xiao and White, Alec and Whitfield, James Daniel and Williamson, Mark J. and Wouters, Sebastian and Yang, Jun and Yu, Jason M. and Zhu, Tianyu and Berkelbach, Timothy C. and Sharma, Sandeep and Sokolov, Alexander Yu. and Chan, Garnet Kin-Lic},
	year = {2020},
	pages = {024109},
}

@article{ziegler_calculation_1977,
	title = {On the calculation of multiplet energies by the hartree-fock-slater method},
	volume = {43},
	number = {3},
	urldate = {2025-07-30},
	journal = {Theoret. Chim. Acta},
	author = {Ziegler, Tom and Rauk, Arvi and Baerends, Evert J.},
	year = {1977},
	pages = {261--271},
}

@article{bagus_singlettriplet_1975,
	title = {Singlet–triplet splittings as obtained from the {X$\alpha$}-scattered wave method: {A} theoretical analysis},
	volume = {9},
	shorttitle = {Singlet–triplet splittings as obtained from the {Xα}-scattered wave method},
	number = {1},
	urldate = {2025-07-30},
	journal = {Int. J. Quantum Chem.},
	author = {Bagus, P. S. and Bennett, B. I.},
	year = {1975},
	pages = {143--148},
}

@article{bene_selfconsistent_1971,
	title = {Self‐{Consistent} {Molecular} {Orbital} {Methods}. {X}. {Molecular} {Orbital} {Studies} of {Excited} {States} with {Minimal} and {Extended} {Basis} {Sets}},
	volume = {55},
	number = {5},
	urldate = {2025-08-26},
	journal = {J. Chem. Phys.},
	author = {Bene, Janet E. Del and Ditchfield, R. and Pople, J. A.},
	year = {1971},
	pages = {2236--2241},
}

@article{schirmer_new_1983,
	title = {New approach to the one-particle {Green}'s function for finite {Fermi} systems},
	volume = {28},
	number = {3},
	urldate = {2025-08-25},
	journal = {Phys. Rev. A},
	author = {Schirmer, J. and Cederbaum, L. S. and Walter, O.},
	year = {1983},
	pages = {1237--1259},
}

@article{schirmer_non-dyson_1998,
	title = {A non-{Dyson} third-order approximation scheme for the electron propagator},
	volume = {109},
	number = {12},
	urldate = {2025-08-26},
	journal = {J. Chem. Phys.},
	author = {Schirmer, J. and Trofimov, A. B. and Stelter, G.},
	year = {1998},
	pages = {4734--4744},
}

@article{dreuw_algebraic_2015,
	title = {The algebraic diagrammatic construction scheme for the polarization propagator for the calculation of excited states},
	volume = {5},
	copyright = {© 2014 John Wiley \& Sons, Ltd.},
	number = {1},
	urldate = {2025-08-26},
	journal = {WIREs Comput. Mol. Sci.},
	author = {Dreuw, Andreas and Wormit, Michael},
	year = {2015},
	pages = {82--95},
}

@article{banerjee_algebraic_2023,
	title = {Algebraic {Diagrammatic} {Construction} {Theory} for {Simulating} {Charged} {Excited} {States} and {Photoelectron} {Spectra}},
	volume = {19},
	number = {11},
	urldate = {2025-08-26},
	journal = {J. Chem. Theory Comput.},
	author = {Banerjee, Samragni and Sokolov, Alexander Yu.},
	year = {2023},
	pages = {3037--3053},
}

@article{hait_orbital_2021,
	title = {Orbital {Optimized} {Density} {Functional} {Theory} for {Electronic} {Excited} {States}},
	volume = {12},
	number = {19},
	urldate = {2025-03-19},
	journal = {J. Phys. Chem. Lett.},
	author = {Hait, Diptarka and Head-Gordon, Martin},
	year = {2021},
	pages = {4517--4529},
}

@article{hait_accurate_2020,
	title = {Accurate prediction of core-level spectra of radicals at density functional theory cost via square gradient minimization and recoupling of mixed configurations},
	volume = {153},
	number = {13},
	urldate = {2025-03-19},
	journal = {J. Chem. Phys.},
	author = {Hait, Diptarka and Haugen, Eric A. and Yang, Zheyue and Oosterbaan, Katherine J. and Leone, Stephen R. and Head-Gordon, Martin},
	year = {2020},
	pages = {134108},
}

@article{ye_-scf_2017,
	title = {{$\sigma$}-{SCF}: {A} direct energy-targeting method to mean-field excited states},
	volume = {147},
	shorttitle = {σ-{SCF}},
	number = {21},
	urldate = {2025-03-19},
	journal = {J. Chem. Phys.},
	author = {Ye, Hong-Zhou and Welborn, Matthew and Ricke, Nathan D. and Van Voorhis, Troy},
	year = {2017},
	pages = {214104},
}

@article{panades-barrueta_accelerating_2023,
	title = {Accelerating {Core}-{Level} {GW} {Calculations} by {Combining} the {Contour} {Deformation} {Approach} with the {Analytic} {Continuation} of {W}},
	volume = {19},
	number = {16},
	urldate = {2025-08-27},
	journal = {J. Chem. Theory Comput.},
	author = {Panadés-Barrueta, Ramón L. and Golze, Dorothea},
	year = {2023},
	pages = {5450--5464},
}

@article{hybertsen_electron_1986,
	title = {Electron correlation in semiconductors and insulators: {Band} gaps and quasiparticle energies},
	volume = {34},
	copyright = {http://link.aps.org/licenses/aps-default-license},
	shorttitle = {Electron correlation in semiconductors and insulators},
	number = {8},
	urldate = {2024-05-08},
	journal = {Phys. Rev. B},
	author = {Hybertsen, Mark S. and Louie, Steven G.},
	year = {1986},
	pages = {5390--5413},
}

@article{blase_first-principles_2011,
	title = {First-principles ${\mathit {{GW}}}$ calculations for fullerenes, porphyrins, phtalocyanine, and other molecules of interest for organic photovoltaic applications},
	volume = {83},
	number = {11},
	urldate = {2025-08-27},
	journal = {Phys. Rev. B},
	author = {Blase, X. and Attaccalite, C. and Olevano, V.},
	year = {2011},
	pages = {115103},
}

@inbook{doi:10.1142/9789812830586,
author = {MARK E. CASIDA},
title = {Time-Dependent Density Functional Response Theory for Molecules},
booktitle = {Recent Advances in Density Functional Methods},
chapter = {},
year = {1995},
pages = {155-192},
}

@article{runge_density-functional_1984,
	title = {Density-{Functional} {Theory} for {Time}-{Dependent} {Systems}},
	volume = {52},
	number = {12},
	urldate = {2025-08-27},
	journal = {Phys. Rev. Lett.},
	author = {Runge, Erich and Gross, E. K. U.},
	year = {1984},
	pages = {997--1000},
}

@article{bauernschmitt_treatment_1996,
	title = {Treatment of electronic excitations within the adiabatic approximation of time dependent density functional theory},
	volume = {256},
	number = {4},
	urldate = {2025-08-27},
	journal = {Chem. Phys. Lett.},
	author = {Bauernschmitt, Rüdiger and Ahlrichs, Reinhart},
	year = {1996},
	pages = {454--464},
}

@article{della_sala_excitation_2003,
	title = {Excitation energies of molecules by time-dependent density functional theory based on effective exact exchange {Kohn}–{Sham} potentials},
	volume = {91},
	copyright = {Copyright © 2003 Wiley Periodicals, Inc.},
	number = {2},
	urldate = {2025-08-27},
	journal = {Int. J. Quantum Chem.},
	author = {Della Sala, Fabio and Görling, Andreas},
	year = {2003},
	pages = {131--138},
}

@article{salpeter_relativistic_1951,
	title = {A {Relativistic} {Equation} for {Bound}-{State} {Problems}},
	volume = {84},
	number = {6},
	urldate = {2025-08-27},
	journal = {Phys. Rev.},
	author = {Salpeter, E. E. and Bethe, H. A.},
	year = {1951},
	pages = {1232--1242},
}

@article{rohlfing_electron-hole_2000,
	title = {Electron-hole excitations and optical spectra from first principles},
	volume = {62},
	number = {8},
	urldate = {2024-09-18},
	journal = {Phys. Rev. B},
	author = {Rohlfing, Michael and Louie, Steven G.},
	year = {2000},
	pages = {4927--4944},
}

@book{martin_interacting_2013,
	address = {Cambridge},
	title = {Interacting {Electrons}: {Theory} and {Computational} {Approaches}},
	isbn = {978-1-139-05080-7},
	shorttitle = {Interacting {Electrons}},
	author = {Martin, Richard M. and Reining, Lucia and Ceperley, David M.},
	year = {2013},
}

@article{onida_electronic_2002,
	title = {Electronic excitations: density-functional versus many-body {Green}'s-function approaches},
	volume = {74},
	shorttitle = {Electronic excitations},
	number = {2},
	urldate = {2025-04-30},
	journal = {Rev. Mod. Phys.},
	author = {Onida, Giovanni and Reining, Lucia and Rubio, Angel},
	year = {2002},
	pages = {601--659},
}

@article{hedin_new_1965,
	title = {New {Method} for {Calculating} the {One}-{Particle} {Green}'s {Function} with {Application} to the {Electron}-{Gas} {Problem}},
	volume = {139},
	number = {3A},
	journal = {Phys. Rev.},
	author = {Hedin, Lars},
	year = {1965},
	pages = {A796},
}

@article{mei_exact_2021,
	title = {Exact {Second}-{Order} {Corrections} and {Accurate} {Quasiparticle} {Energy} {Calculations} in {Density} {Functional} {Theory}},
	volume = {12},
	number = {30},
	urldate = {2024-02-28},
	journal = {J. Phys. Chem. Lett.},
	author = {Mei, Yuncai and Chen, Zehua and Yang, Weitao},
	year = {2021},
	pages = {7236--7244},
}

@article{dunning_gaussian_1989,
	title = {Gaussian {Basis} {Sets} for {Use} in {Correlated} {Molecular} {Calculations}. {I}. {The} {Atoms} {Boron} through {Neon} and {Hydrogen}},
	volume = {90},
	journal = {J. Chem. Phys.},
	author = {Dunning, Thom H},
	year = {1989},
	pages = {1007--1023},
}

@article{kendall_electron_1992,
	title = {Electron {Affinities} of the {First}-{Row} {Atoms} {Revisited}. {Systematic} {Basis} {Sets} and {Wave} {Functions}},
	volume = {96},
	journal = {J. Chem. Phys.},
	author = {Kendall, Rick A and Dunning, Thom H and Harrison, Robert J},
	year = {1992},
	pages = {6796--6806},
}

@article{schafer_fully_1992,
	title = {Fully optimized contracted {Gaussian} basis sets for atoms {Li} to {Kr}},
	volume = {97},
	number = {4},
	urldate = {2025-08-27},
	journal = {J. Chem. Phys.},
	author = {Schäfer, Ansgar and Horn, Hans and Ahlrichs, Reinhart},
	year = {1992},
	pages = {2571--2577},
}

@article{lee_development_1988,
	title = {Development of the {Colle}-{Salvetti} correlation-energy formula into a functional of the electron density},
	volume = {37},
	number = {2},
	urldate = {2025-08-27},
	journal = {Phys. Rev. B},
	author = {Lee, Chengteh and Yang, Weitao and Parr, Robert G.},
	year = {1988},
	pages = {785--789},
}

@article{perdew_generalized_1996,
	title = {Generalized {Gradient} {Approximation} {Made} {Simple}},
	volume = {77},
	number = {18},
	urldate = {2025-08-27},
	journal = {Phys. Rev. Lett.},
	author = {Perdew, John P. and Burke, Kieron and Ernzerhof, Matthias},
	year = {1996},
	pages = {3865--3868},
}

@article{seidu_applications_2015,
	title = {Applications of {Time}-{Dependent} and {Time}-{Independent} {Density} {Functional} {Theory} to {Electronic} {Transitions} in {Tetrahedral} d0 {Metal} {Oxides}},
	volume = {11},
	number = {9},
	urldate = {2026-02-04},
	journal = {J. Chem. Theory Comput.},
	author = {Seidu, Issaka and Krykunov, Mykhaylo and Ziegler, Tom},
	year = {2015},
	pages = {4041--4053},
}

@article{vandaele_scf_2022,
	title = {The {$\Delta$}{SCF} method for non-adiabatic dynamics of systems in the liquid phase},
	volume = {156},
	number = {13},
	urldate = {2025-03-10},
	journal = {J. Chem. Phys.},
	author = {Vandaele, Eva and Mali, Momir and Luber, Sandra},
	year = {2022},
	pages = {130901},
}

@incollection{SLATER19721,
title = {Statistical Exchange-Correlation in the Self-Consistent Field},
editor = {Per-Olov Löwdin},
series = {Advances in Quantum Chemistry},
volume = {6},
pages = {1-92},
year = {1972},
author = {John C. Slater},
}

@article{slater_statistical_1970,
	title = {Statistical exchange and the total energy of a crystal},
	volume = {5},
	number = {S4},
	urldate = {2026-01-22},
	journal = {Int. J. Quantum Chem.},
	author = {Slater, John C. and Wood, John H.},
	year = {1970},
	pages = {3--34},
}

@article{cheng_rydberg_2008,
	title = {Rydberg energies using excited state density functional theory},
	volume = {129},
	number = {12},
	urldate = {2026-01-23},
	journal = {J. Chem. Phys.},
	author = {Cheng, Chiao-Lun and Wu, Qin and Van Voorhis, Troy},
	year = {2008},
	pages = {124112},
}

@article{RevModPhys.61.689,
  title = {The density functional formalism, its applications and prospects},
  author = {Jones, R. O. and Gunnarsson, O.},
  journal = {Rev. Mod. Phys.},
  volume = {61},
  issue = {3},
  pages = {689--746},
  numpages = {0},
  year = {1989},
}

@article{doi:10.1021/acs.jctc.6b01161,
author = {Liu, Junzi and Zhang, Yong and Bao, Peng and Yi, Yuanping},
title = {Evaluating Electronic Couplings for Excited State Charge Transfer Based on Maximum Occupation Method {$\Delta$}SCF Quasi-Adiabatic States},
journal = {J. Chem. Theory Comput.},
volume = {13},
number = {2},
pages = {843-851},
year = {2017},
}

@article{perdew_self-interaction_1981,
	title = {Self-interaction correction to density-functional approximations for many-electron systems},
	volume = {23},
	number = {10},
	urldate = {2026-01-21},
	journal = {Phys. Rev. B},
	author = {Perdew, J. P. and Zunger, Alex},
	year = {1981},
	pages = {5048--5079},
}

@article{PhysRevLett.49.1691,
  title = {Density-Functional Theory for Fractional Particle Number: Derivative Discontinuities of the Energy},
  author = {Perdew, John P. and Parr, Robert G. and Levy, Mel and Balduz, Jose L.},
  journal = {Phys. Rev. Lett.},
  volume = {49},
  issue = {23},
  pages = {1691--1694},
  numpages = {0},
  year = {1982},
}

@article{PhysRevLett.84.5172,
  title = {Degenerate Ground States and a Fractional Number of Electrons in Density and Reduced Density Matrix Functional Theory},
  author = {Yang, Weitao and Zhang, Yingkai and Ayers, Paul W.},
  journal = {Phys. Rev. Lett.},
  volume = {84},
  issue = {22},
  pages = {5172--5175},
  numpages = {0},
  year = {2000},
}

@article{zhang_perspective_2000,
	title = {Perspective on “{Density}-functional theory for fractional particle number: derivative discontinuities of the energy”},
	volume = {103},
	shorttitle = {Perspective on “{Density}-functional theory for fractional particle number},
	number = {3},
	urldate = {2026-02-04},
	journal = {Theor Chem Acc},
	author = {Zhang, Yingkai and Yang, Weitao},
	year = {2000},
	pages = {346--348},
}

@article{PhysRevA.76.040501,
  title = {Exchange and correlation in open systems of fluctuating electron number},
  author = {Perdew, John P. and Ruzsinszky, Adrienn and Csonka, G\'abor I. and Vydrov, Oleg A. and Scuseria, Gustavo E. and Staroverov, Viktor N. and Tao, Jianmin},
  journal = {Phys. Rev. A},
  volume = {76},
  issue = {4},
  pages = {040501},
  numpages = {4},
  year = {2007},
}

@article{zhou_all-electron_2025,
	title = {All-{Electron} {BSE}@{GW} {Method} with {Numeric} {Atom}-{Centered} {Orbitals} for {Extended} {Periodic} {Systems}},
	volume = {21},
	number = {1},
	urldate = {2025-07-08},
	journal = {J. Chem. Theory Comput.},
	author = {Zhou, Ruiyi and Yao, Yi and Blum, Volker and Ren, Xinguo and Kanai, Yosuke},
	year = {2025},
	pages = {291--306},
}

@article{li_localized_2018,
	title = {Localized orbital scaling correction for systematic elimination of delocalization error in density functional approximations},
	volume = {5},
	copyright = {http://creativecommons.org/licenses/by/4.0/},
	number = {2},
	urldate = {2024-03-29},
	journal = {Natl. Sci. Rev.},
	author = {Li, Chen and Zheng, Xiao and Su, Neil Qiang and Yang, Weitao},
	year = {2018},
	pages = {203--215},
}

@article{yu_accurate_2025,
	title = {Accurate {Prediction} of {Core}-{Level} {Binding} {Energies} from {Ground}-{State} {Density} {Functional} {Calculations}: {The} {Importance} of {Localization} and {Screening}},
	volume = {16},
	shorttitle = {Accurate {Prediction} of {Core}-{Level} {Binding} {Energies} from {Ground}-{State} {Density} {Functional} {Calculations}},
	number = {10},
	urldate = {2025-03-17},
	journal = {J. Phys. Chem. Lett.},
	author = {Yu, Jincheng and Mei, Yuncai and Chen, Zehua and Fan, Yichen and Yang, Weitao},
	year = {2025},
	pages = {2492--2500},
}

@article{su_preserving_2020,
	title = {Preserving {Symmetry} and {Degeneracy} in the {Localized} {Orbital} {Scaling} {Correction} {Approach}},
	volume = {11},
	number = {4},
	urldate = {2024-03-22},
	journal = {J. Phys. Chem. Lett.},
	author = {Su, Neil Qiang and Mahler, Aaron and Yang, Weitao},
	year = {2020},
	pages = {1528--1535},
}

@article{li_localized_2025,
	title = {Localized {Orbital} {Scaling} {Correction} to {Linear}-{Response} {Time}-{Dependent} {Density} {Functional} {Approximations}},
	urldate = {2025-06-06},
	journal = {J. Chem. Theory Comput.},
	author = {Li, Ye and Li, Chen},
	year = {2025},
}

@book{ullrich_time-dependent_2012,
	address = {Oxford, New York},
	series = {Oxford {Graduate} {Texts}},
	title = {Time-{Dependent} {Density}-{Functional} {Theory}: {Concepts} and {Applications}},
	isbn = {978-0-19-956302-9},
	shorttitle = {Time-{Dependent} {Density}-{Functional} {Theory}},
	publisher = {Oxford University Press},
	author = {Ullrich, Carsten},
	year = {2012},
}

@article{zhu_all-electron_2021,
	title = {All-{Electron} {Gaussian}-{Based} {G0W0} for {Valence} and {Core} {Excitation} {Energies} of {Periodic} {Systems}},
	volume = {17},
	number = {2},
	urldate = {2026-02-04},
	journal = {J. Chem. Theory Comput.},
	author = {Zhu, Tianyu and Chan, Garnet Kin-Lic},
	year = {2021},
	pages = {727--741},
}

@article{liu_all-electron_2020,
	title = {All-electron \textit{ab initio} {Bethe}-{Salpeter} equation approach to neutral excitations in molecules with numeric atom-centered orbitals},
	volume = {152},
	copyright = {https://publishing.aip.org/authors/rights-and-permissions},
	number = {4},
	urldate = {2024-09-17},
	journal = {J. Chem. Phys.},
	author = {Liu, Chi and Kloppenburg, Jan and Yao, Yi and Ren, Xinguo and Appel, Heiko and Kanai, Yosuke and Blum, Volker},
	year = {2020},
}

@article{blum_ab_2009,
	title = {\textit{{Ab} initio} molecular simulations with numeric atom-centered orbitals},
	volume = {180},
	number = {11},
	urldate = {2026-02-04},
	journal = {Comput. Phys. Commun.},
	author = {Blum, Volker and Gehrke, Ralf and Hanke, Felix and Havu, Paula and Havu, Ville and Ren, Xinguo and Reuter, Karsten and Scheffler, Matthias},
	year = {2009},
	pages = {2175--2196},
}

@article{li_piecewise_2017,
	title = {On the piecewise convex or concave nature of ground state energy as a function of fractional number of electrons for approximate density functionals},
	volume = {146},
	number = {7},
	urldate = {2026-02-05},
	journal = {J. Chem. Phys.},
	author = {Li, Chen and Yang, Weitao},
	year = {2017},
	pages = {074107},
}

@article{bhattacharya_bsegw_2024,
	title = {{BSE}@{GW} {Prediction} of {Charge} {Transfer} {Exciton} in {Molecular} {Complexes}: {Assessment} of {Self}-{Energy} and {Exchange}-{Correlation} {Dependence}},
	volume = {128},
	shorttitle = {{BSE}@{GW} {Prediction} of {Charge} {Transfer} {Exciton} in {Molecular} {Complexes}},
	number = {29},
	urldate = {2026-02-05},
	journal = {J. Phys. Chem. A},
	author = {Bhattacharya, Sampreeti and Li, Jiachen and Yang, Weitao and Kanai, Yosuke},
	year = {2024},
	pages = {6072--6083},
}

@article{jiang_gw_2016,
	title = {${GW}$ with linearized augmented plane waves extended by high-energy local orbitals},
	volume = {93},
	number = {11},
	urldate = {2026-02-05},
	journal = {Phys. Rev. B},
	author = {Jiang, Hong and Blaha, Peter},
	year = {2016},
	pages = {115203},
}

@article{wang_time-dependent_2004,
	title = {Time-dependent density functional theory based on a noncollinear formulation of the exchange-correlation potential},
	volume = {121},
	number = {24},
	urldate = {2025-09-10},
	journal = {J. Chem. Phys.},
	author = {Wang, Fan and Ziegler, Tom},
	year = {2004},
	pages = {12191--12196},
}

\end{document}